\documentclass[a4paper,twocolumn,11pt,accepted=2022-05-16]{quantumarticle}
\pdfoutput=1
\usepackage[utf8]{inputenc}
\usepackage[english]{babel}
\usepackage[T1]{fontenc}
\usepackage{amsmath}
\usepackage{hyperref}

\usepackage[numbers, sort & compress]{natbib}

\usepackage{tikz}
\usepackage{lipsum}

\begin{document}

\title{
Analytical framework for non-equilibrium phase transition to Bose--Einstein condensate}

\author{V. Yu. Shishkov}
\email{vladislavmipt@gmail.com}
\affiliation{
 Dukhov Research Institute of Automatics (VNIIA), 22 Sushchevskaya, Moscow 127055, Russia;
}
\orcid{0000-0002-2445-2701}
\affiliation{
 Moscow Institute of Physics and Technology, 9 Institutskiy pereulok, Dolgoprudny 141700, Moscow region, Russia;
}
\affiliation{
 Center for Photonics and Quantum Materials, Skolkovo Institute of Science and Technology, Moscow, Russian Federation 
}
\affiliation{
 Laboratories for Hybrid Photonics, Skolkovo Institute of Science and Technology, Moscow, Russian Federation
}

\author{E. S. Andrianov}
\affiliation{
 Dukhov Research Institute of Automatics (VNIIA), 22 Sushchevskaya, Moscow 127055, Russia;
}
\affiliation{
 Moscow Institute of Physics and Technology, 9 Institutskiy pereulok, Dolgoprudny 141700, Moscow region, Russia;
}
\affiliation{
 Center for Photonics and Quantum Materials, Skolkovo Institute of Science and Technology, Moscow, Russian Federation 
}
\affiliation{
 Laboratories for Hybrid Photonics, Skolkovo Institute of Science and Technology, Moscow, Russian Federation
}

\author{Yu. E. Lozovik}
\affiliation{
 Dukhov Research Institute of Automatics (VNIIA), 22 Sushchevskaya, Moscow 127055, Russia;
}
\affiliation{
 Moscow Institute of Electronics and Mathematics, National Research University Higher School of Economics, 101000 Moscow, Russia;
}
\affiliation{
 Institute for Spectroscopy RAS, 5 Fizicheskaya, Troitsk 142190, Russia;
}

\maketitle

\begin{abstract}
The theoretical description of non-equilibrium Bose--Einstein condensate (BEC) is one of the main challenges in modern statistical physics and kinetics.
The non-equilibrium nature of BEC makes it impossible to employ the well-established formalism of statistical mechanics.
We develop a framework for the analytical description of a non-equilibrium phase transition to BEC that, in contrast to previously developed approaches, takes into account the infinite number of continuously distributed states.
We consider the limit of fast thermalization and obtain an analytical expression for the full density matrix of a non-equilibrium ideal BEC which also covers the equilibrium case.
For the particular cases of 2D and 3D, we investigate the non-equilibrium formation of BEC by finding the temperature dependence of the ground state occupation and second-order coherence function.
We show that for a given pumping rate, the macroscopic occupation of the ground state and buildup of coherence may occur at different temperatures.
Moreover, the buildup of coherence strongly depends on the pumping scheme.
We also investigate the condensate linewidth and show that the Schawlow–Townes law holds for BEC in 3D and does not hold for BEC in 2D.

\end{abstract}

\section{Introduction} \label{sec:introduction}

Non-equilibrium Bose--Einstein condensates (BECs) working at room temperature~\cite{plumhof2014room, zasedatelev2019room} are systems where open dissipative quantum-mechanical nature meets many-body collective dynamics.
Besides its fundamental theoretical interest, BEC can be used in sub-picosecond switching at the fundamental quantum limit~\cite{zasedatelev2021single}, all-optical manipulation, and other low-energy optoelectronic applications~\cite{sanvitto2016road, keeling2020bose}.
Among the most promising systems for the realization of BEC are exciton-polaritons, the hybrid quasiparticles formed by the strong light--matter interaction of photons in a cavity and excitons,~\cite{deng2003polariton, kasprzak2006bose, combescot2015excitons, byrnes2014exciton, wertz2010spontaneous, balili2007bose, estrecho2018single, sun2017bose, deng2010exciton, klaas2018photon, imamog1996nonequilibrium, keeling2020bose, wei2019low, zasedatelev2019room, plumhof2014room}. 
The advantage of the exciton-polaritons is their small effective mass and low density of states~\cite{byrnes2014exciton, wertz2010spontaneous, deng2010exciton, imamog1996nonequilibrium}, that provides a critical temperature of hundreds of kelvin~\cite{deng2003polariton, kasprzak2006bose, plumhof2014room}.

In many BEC realizations, the thermalization rate of the polaritons is much faster than their dissipation rate and pumping rate.
Namely, for long-lived polariton and photon systems, thermalization occurs at pumping rates significantly lower than the condensation threshold~\cite{sun2017bose, weill2019bose}.
The relatively high thermal energy in polariton systems at room temperature facilitates phonon- and vibron-assisted polariton relaxation, which, overall, results in efficient polariton thermalization.  
A fast thermalization in an organic polariton system has been revealed with an extreme rate on the order of 200~fs~\cite{hakala2018bose, vakevainen2020sub}.

Previous studies of BEC in these systems largely rely on the semiclassical Maxwell--Boltzmann equations describing the average population of exciton-polariton states~\cite{malpuech2002room, banyai2002real, cao2004condensation, doan2008coherence, tassone1997bottleneck}, other mean-field theories~\cite{kirton2013nonequilibrium, kirton2015thermalization, strashko2018organic}, and an approach beyond mean-field theory~\cite{arnardottir2020multimode} that takes into account the higher-order correlations.
However, in such treatments, information about the correlations between different polariton states is lost.
In many cases, the driven dissipative Gross--Pitaevskii equation with noise successfully describes the dynamics of BEC~\cite{carusotto2013quantum}. 
However, the Gross--Pitaevskii equation works well only when temperature is well below the condensation temperature~\cite{carusotto2013quantum}. 
This means that it cannot accurately describe the crossover and buildup of coherence.

The dynamics of the polaritons can be described by the corresponding master equation for the density matrix~\cite{kavokin2017microcavities, sanvitto2012exciton}, with transition rates between quantum states in a configuration space rather than between populations as in semiclassical Maxwell--Boltzmann equations.
The density matrix enables finding correlations of any order, but the master equation become hard to solve numerically for large systems. 
The analytical steady state solution for the density matrix for two-oscillators system were obtained in~\cite{laussy2004spontaneousPSSC, laussy2004spontaneousPRL} and generalized for arbitrary degeneracy of the excited state in~\cite{sanvitto2012exciton}.
In these works, the authors discussed how the number-conserving scattering processes allow to obtain this analytical solution. 
Relying on similar ideas, recently, it has been shown that, in the case of fast thermalization of polaritons, the density matrix can be obtained analytically for arbitrary finite sets of polariton states~\cite{shishkov2021exact}.
However, due to the finiteness of the states, this approach is limited in its ability to analytically investigate the influence of the polaritons' dispersion and dimension on the condensation threshold, buildup of coherence, long-rage spatial correlations, and linewidth. 
Thus, a comprehensive analysis of BEC requires the extension of the approach~\cite{shishkov2021exact} to an infinite number of states.

In the present paper, we develop a formalism which extends the advantages of the exactly solvable Lindblad master equation for non-equilibrium BEC in the fast thermalization limit~\cite{shishkov2021exact} to the systems with continuously distributed states. 
This developed approach consists of two essential steps. 
At the first step, it is assumed that in each sector of Fock space with fixed particle number $N$ distributed in the whole system, the density matrix $\hat\rho_N$ is in thermal equilibrium. 
This assumption is fulfilled if thermalizing processes are much faster than processes that change the particle number (external pumping and dissipation)~\cite{shishkov2021exact}. 
In this case, the total system density matrix can be written as a time-dependent sum of thermal, canonical density matrices in each particle number sector. 
Thus, the problem is effectively reduced to calculating all thermal density matrices for arbitrary, but fixed particle number.
At the second step, we develop a recursive method for calculating all equilibrium density matrices with fixed particle number.
As a result, we obtain the full density matrix of all the states. 
We show that, for a non-equilibrium BEC in the fast thermalization limit, the condition for the macroscopic occupation of the ground state is the same as in the full thermal equilibrium case.
However, the buildup of coherence in a non-equilibrium BEC depends not only on the number of polaritons and the environmental temperature, but also on the pumping scheme.
An increase in the pumping rate of the ground state lead to the macroscopic occupation of the ground state, but destroys the coherence of a non-equilibrium BEC; on the contrary, an increase in the pumping rate of the excited states results in both the macroscopic occupation of the ground state and the buildup of coherence.
We also show that the linewidth of the condensate narrows with an increase of the macroscopic occupation of the ground state, but in 2D this narrowing does not follow the Schawlow--Townes equation.

\section{Description of the model}
The strong light--matter interaction of a cavity electric field with the dipole moments of an optical transition of molecules gives rise to new eigenstates, namely, lower and upper polariton branches.
Due to dissipation processes, it is necessary to compensate for the losses, which can be done by an external pump.
Usually, an external pump excites polaritons in the upper polariton branch and excitons which are not coupled with the cavity modes~\cite{zasedatelev2019room, byrnes2014exciton}.
The polaritons of the upper branch and excitons~\cite{zasedatelev2019room, byrnes2014exciton} scatter the energy supplied by the pump into the lower polariton branch forming BEC at high enough excitation density.
There are several mechanisms allowing such energy relaxation in polariton systems. 
Scattering by electrons~\cite{lagoudakis2003electron}, high-energy optical phonons~\cite{maragkou2010longitudinal}, and vibrons in organic materials~\cite{zasedatelev2019room,coles2011vibrationally} are among them. 
Since non-equilibrium BEC occurs mostly in the lower polariton branch, below we consider the dynamics of the lower polariton branch only.
We describe the energy transfer from the upper polaritons or excitons to the lower polaritons by an effective incoherent pumping.
Such an approximation is reasonable for the majority of experimental realizations~\cite{kasprzak2006bose,plumhof2014room,sun2017bose}

We consider a continuous system of polariton states with the frequency $\omega_{\bf k}$ for a wave vector ${\bf k}$ in the general case of $D$ dimensional space.
We suppose that the polaritons with wave vector ${\bf k}$ of the lower polariton branch are described by bosonic creation $\hat a_{\bf k}^\dag $ and annihilation $\hat a_{\bf k}$ operators~\cite{kavokin2017microcavities, sanvitto2012exciton} which obey the commutation relation $\left[ {{{\hat a}_{\bf k}},\hat a_{\bf q}^\dag } \right] = {\delta ({\bf k} - {\bf q})}$. 
In this case, the Hamiltonian of the polaritons takes the form 
\begin{equation}\label{Hamiltonian}
{\hat H_{{\rm{LP}}}} = \sum_{\bf k} {{\omega_{\bf k}}\hat a_{\bf k}^\dag {{\hat a}_{\bf k}}} ,
\end{equation}
where the state with ${\bf k}={\bf 0}$ is the ground state. 
Hereafter, we assume that $\hbar=1$.
We consider the ideal polaritons, i.e. there is no interaction between polaritons in the Hermitian part.

We describe the dynamics of the polaritons in the lower polariton branch through the density matrix $\hat \rho$ and consider the relaxation and pumping processes in the Born--Markov approximation~\cite{breuer2002theory}. 
In this approximation, the density matrix is governed by the master equation in the Lindblad form. 
The dissipation of the lower polaritons is governed by the Lindblad superoperator~\cite{kavokin2017microcavities} 
\begin{equation}\label{Lindblad_diss}
{L_{{\rm{diss}}}}\left( {\hat \rho } \right) = \sum_{\bf k} {{\gamma_{\bf k}}\left( {{{\hat a}_{\bf k}}\hat \rho \hat a_{\bf k}^\dag  - \frac{1}{2}\hat \rho \hat a_{\bf k}^\dag {{\hat a}_{\bf k}} - \frac{1}{2}\hat a_{\bf k}^\dag {{\hat a}_{\bf k}}\hat \rho } \right)} ,
\end{equation}
where ${\gamma_{\bf k}}$ is the dissipation rate of the state with wave vector ${\bf k}$. 
The effective incoherent pumping of the lower polaritons can be described by the Lindblad superoperator
\begin{multline}\label{Lindblad_pump}
{L_{{\rm{pump}}}}\left( {\hat \rho } \right) = \\
	\sum_{\bf k} {{\kappa_{\bf k}}\left( {{{\hat a}_{\bf k}}\hat \rho \hat a_{\bf k}^\dag  - \frac{1}{2}\hat \rho \hat a_{\bf k}^\dag {{\hat a}_{\bf k}} - \frac{1}{2}\hat a_{\bf k}^\dag {{\hat a}_{\bf k}}\hat \rho } \right)}  \\
+ \sum_{\bf k} {{\kappa_{\bf k}}\left( {\hat a_{\bf k}^\dag \hat \rho {{\hat a}_{\bf k}} - \frac{1}{2}\hat \rho {{\hat a}_{\bf k}}\hat a_{\bf k}^\dag - \frac{1}{2}{{\hat a}_{\bf k}}\hat a_{\bf k}^\dag \hat \rho } \right)}, 
\end{multline}
where ${\kappa_{\bf k}}$ is the pumping rate of the state with wave vector ${\bf k}$. 
The Lindblad operator $L_{\rm pump}$ leads to the following dynamics of the average polariton population in the state with the wave vector $\bf k$~\cite{shishkov2021exact}:
$
\left( d\langle \hat a_{\bf k}^\dag \hat a_{\bf k} \rangle/dt \right)_{\rm pump} = \kappa_{\bf k}
$ and
$
\left( d\langle \hat a_{\bf k} \rangle/dt \right)_{\rm pump} = 0
$.
Therefore, $L_{\rm pump}$ leads to the excitation of a certain number of polaritons per unit time in the
corresponding state without affecting the phase of the state.

Thermalization of the lower polaritons may occur due to different physical processes depending on the system.
For example, in organic polariton systems, thermalization occurs due to their nonlinear interaction with low frequency vibrations~\cite{strashko2018organic, litinskaya2004fast, mazza2009organic, bittner2012estimating, cwik2014polariton, somaschi2011ultrafast, ramezani2018nonlinear}. 
For polariton states in inorganic semiconductors, thermalization predominantly goes through interactions with acoustic phonons or free charges~\cite{kavokin2017microcavities}.
The thermalization of the polaritons can also occur due to polariton--polariton scattering~\cite{deng2010exciton, savvidis2000angle}.
However, regardless of the mechanism, the thermalization can be described through the Lindblad superoperator~\cite{kavokin2017microcavities, sanvitto2012exciton}
\begin{multline}\label{Lindblad_therm}
{L_{{\rm{therm}}}}\left( {\hat \rho } \right) = \sum_{\bf k} \sum_{{\bf q}\neq {\bf k}} {\Gamma _{\bf kq}}\times \\
 {\left( {{{\hat a}_{\bf k}}\hat a_{\bf q}^\dag \hat \rho {{\hat a}_{\bf q}}\hat a_{\bf k}^\dag  - \frac{1}{2}\hat \rho {{\hat a}_{\bf q}}\hat a_{\bf k}^\dag {{\hat a}_{\bf k}}\hat a_{\bf q}^\dag  - \frac{1}{2}{{\hat a}_{\bf q}}\hat a_{\bf k}^\dag {{\hat a}_{\bf k}}\hat a_{\bf q}^\dag \hat \rho } \right)} ,
\end{multline}
where ${\Gamma_{\bf kq}}$ is the transition rate from the state with the wave vector $\bf k$ to the state with the wave vector $\bf q$. 
The thermalization rates ${\Gamma_{\bf kq}}$ obey the Kubo--Martin--Schwinger relation ${\Gamma_{\bf kq}}/{\Gamma_{\bf qk}} = \exp \left( {\left( {{\omega_{\bf k}} - {\omega_{\bf q}}} \right)/T} \right)$, where $T$ is the temperature of intermolecular vibrations of the organic dyes or the temperature of the phonons in the semiconductors~\cite{kosloff2013quantum}. 
Hereafter, we assume the Boltzmann constant to be equal to unity. 

Thus, the complete master equation for the density matrix of the polaritons of the lower polariton branch $\hat \rho$ has the form
\begin{multline}\label{Master_equation}
{\frac{\partial \hat \rho(t) }{\partial t}} = {-i}\left[ {\hat \rho (t),{{\hat H}_{{\rm{LP}}}}} \right] + \\
+ {L_{{\rm{diss}}}}\left( {\hat \rho (t)} \right) + {L_{{\rm{pump}}}}\left( {\hat \rho (t)} \right) + {L_{{\rm{therm}}}}\left( {\hat \rho (t)} \right).
\end{multline}

\section{Density matrix in the fast thermalization limit} \label{sec:density_matrix}
Further, we assume that the thermalization is the fastest process in the system, i.e., thermalization rates are much higher than the dissipation and pumping rates, ${\Gamma_{\bf 0k}} (1 + \langle \hat a^\dag_{\bf 0}\hat a_{\bf 0}\rangle) \gg {\kappa_{\bf k}},{\gamma_{\bf k}}$. 
In such a case, at times $t \ll \gamma_{\bf k}^{ - 1},\kappa _{\bf k}^{ - 1}$, only thermalization affects the system dynamics.
As a result, the density matrix obeys the approximate differential equation 
\begin{equation}\label{Diff_Fast_thermalization}
\frac{{\partial\hat \rho (t)}}{{\partial t}} = -i\left[ {\hat \rho (t),{{\hat H}_{{\rm{LP}}}}} \right] +{L_{{\rm{therm}}}}\left( {\hat \rho (t)} \right).
\end{equation}
After the time ${\Gamma _{\bf 0k}}^{-1}(1+\langle \hat a^\dag_{\bf 0}\hat a_{\bf 0}\rangle)^{ - 1} \ll t \ll \gamma_{\bf k}^{ - 1},\kappa_{\bf k}^{ - 1}$, the system reaches its quasistationary state determined by
\begin{equation}\label{Fast_thermalization}
-i\left[ {\hat \rho (t),{{\hat H}_{{\rm{LP}}}}} \right] +{L_{{\rm{therm}}}}\left( {\hat \rho (t)} \right) = 0.
\end{equation}
The relaxation operator ${L_{{\rm{therm}}}}\left( {\hat \rho } \right)$, defined by the expression~\eqref{Lindblad_therm}, conserves the total number of lower polaritons~\cite{kavokin2017microcavities}. 
Thus, in the limit ${\kappa_{\bf k}}/{\Gamma _{\bf 0k}}(1+\langle \hat a^\dag_{\bf 0}\hat a_{\bf 0}\rangle) \to 0$ and ${\gamma_{\bf k}}/{\Gamma _{\bf 0k}}(1+\langle \hat a^\dag_{\bf 0}\hat a_{\bf 0}\rangle) \to 0$, the system reaches a quasistationary state governed by Eq.~\eqref{Fast_thermalization} at instantaneous time conserving the total number of polaritons.
Applying the theory developed in~\cite{shishkov2018zeroth}, we can obtain the general form of the density matrix of this quasistationary state.
Because the relaxation operator ${L_{{\rm{therm}}}}\left( {\hat \rho } \right)$ conserves the total number of lower polaritons~\cite{kavokin2017microcavities}, the operator of total number of lower polaritons $\sum_{\bf k}\hat a^\dag_{\bf k}\hat a_{\bf k}$ is an integral of motion for the thermalization process.
According to~\cite{shishkov2018zeroth}, the presence of this integral of motion implies that the system has invariant subspaces with the total number of polaritons equal to $\sum_{\bf k}{n_{\bf k}} = N$.
Being in an invariant subspace at the initial moment in time, the system stays in this invariant subspace in the subsequent moments.
In each invariant subspace with $N$ polaritons, the Gibbs distribution with a temperature $T$ is established~\cite{shishkov2018zeroth}.

The most general form of density matrix satisfying Eq.~\eqref{Fast_thermalization} is the weighted sum of thermalized distributions in each of the invariant subspaces: 
\begin{equation} \label{thermolized_dm}
\hat\rho(t) =
\sum_{N=0}^{+\infty} P_N(t)\hat \rho_N,
\end{equation}
where the summation goes over the total number of polaritons. 
Here $P_N(t)$ is the probability of the system to be in a state with exactly $N$ polaritons distributed over the system as a whole.
These probabilities depend on the time due to the dissipation and pumping processes. 
${\hat \rho}_N$ is the thermalized density matrix of the states with exactly $N$ polaritons, which forms an invariant subspace~\cite{shishkov2021exact}.
This means that $\hat \rho_{N}$ can be represented as 
\begin{equation} \label{thermolized_dm_N}
	{\hat \rho _N} = \frac{1}{{{Z_N}}}\sum\limits_{{\rm config}:\sum\limits_{\bf{k}} {{n_{\bf{k}}}}  = N} {{e^{ - \sum\limits_{\bf{k}} {{n_{\bf{k}}}} ({\omega _{\bf{k}}} - {\omega _{\bf{0}}})/T}}} {\hat R_{\rm{config}}},
\end{equation}
where $Z_N$ is the partition function of the states with exactly $N$ polaritons, and ${\hat R_{\rm{config}}}$ is a diagonal density matrix corresponding to the configuration of the polaritons $\{n_{\bf k}\}$.
This density matrix obeys ${ {\hat a^\dag_{\bf k}}{\hat a_{\bf k}} } {\hat R_{{\rm config}}} = n_{\bf k}{\hat R}_{ \rm{config} }$, $\sum_{\bf{k}} {{n_{\bf{k}}}}  = N$.
We note, that according to~\cite{shishkov2018zeroth} the matrix~\eqref{thermolized_dm_N} is diagonal.
The consideration of diagonal elements is enough to investigate many important properties of BEC such as stationary density matrix, occupation numbers, entropy, coherence, however, we do take into account the dynamics of non-diagonal elements while finding spectral linewidth in Section~\ref{sec:condensation}.

The partition function $Z_N$ is defined by the normalization condition $\rm{Tr} ( {\hat R}_{ \rm{config} } ) = 1$, thus
\begin{equation} \label{partition_function_def}
	Z_N=\sum_{ {\rm config} : \sum_{\bf k} n_{\bf k}=N } 
e^{- \sum_{\bf k} n_{\bf k}(\omega_{\bf k}-\omega_{\bf 0})/{T}}.
\end{equation}
When there are no polaritons in the system, Eq.~\eqref{thermolized_dm_N} implies ${\hat \rho}_{N=0}=Z_0^{-1}| 0 \rangle \langle 0 |$.
Therefore, 
\begin{equation} \label{partition_function_0}
	Z_0=1.
\end{equation}

The density matrix~\eqref{thermolized_dm} allows to calculate all the single-time averages.
In particular, it allows calculating the averages $\langle \hat n_{{\bf k}_1}^{m_1}~...~\hat n_{{\bf k}_M}^{m_M} \rangle$:
\begin{multline} \label{average}
\langle \hat n_{{\bf k}_1}^{m_1}~...~\hat n_{{\bf k}_M}^{m_M} \rangle =\\
\left(-T\right)^{\sum_{j=1}^M m_j} 
\sum_{N=0}^{+\infty}  \frac{P_N(t)}{Z_N} \frac{{\partial^{m_1}}~...~{\partial^{m_M}} Z_N}{ {\partial \omega_{{\bf k}_1}^{m_1}}~...~{\partial \omega_{{\bf k}_M}^{m_M}} }.
\end{multline}
The derivative of $Z_N$ with respect to $\omega_{\bf k}$ is (see Appendix~\ref{sec:appendix_a})
\begin{equation} \label{derivative_Z_N}
\frac{\partial Z_N}{\partial \omega_{\bf k}} = \frac{1}{T} \left[Z_N - \sum_{n=0}^N Z_{N-n} e^{-n(\omega_{\bf k}-\omega_{\bf 0})/{T}}   \right].
\end{equation}
Note that $\partial Z_N/\partial \omega_{\bf 0}$ must be calculated as $\lim_{{\bf k}\to{\bf 0}}\partial Z_N/\partial \omega_{\bf k}$.

\section{Partition function $Z_N$} \label{sec:partition_function}
The aim of paper is to study the process of formation of BEC and its properties in the continuous limit. 
This limit covers not only the system occupying the infinite volume, but also the finite volume as well.
Indeed, the states may be treated as a continuum for systems of finite size when the volume is so large that the energy spacing between the quantized states is small compared to $T$~\cite{berman2008theory}.
However, even if this condition is not satisfied, the analysis presented below may be useful for qualitative analysis of BEC. 

We consider the continuous limit of the partition function~\eqref{partition_function_def} and replace the sum over wave vectors~$\bf k$ by the integral over frequencies $\omega$:
\begin{equation} \label{continuous_limit}
\sum_{\bf k}~... = \int d\omega \left( 1\cdot\delta(\omega-\omega_{\bf 0}) + \nu(\omega) \right)~...
\end{equation}
The density of states $\nu(\omega)$ is  
\begin{equation} \label{density_of_states}
\nu(\omega) = \frac{V}{(2\pi)^D}\left(\frac{d^D {\bf k}}{d\omega_{\bf k}}\right)_{\omega_{\bf k}=\omega},
\end{equation}
where $V$ is the volume in 3D or the area in 2D occupied by polaritons.
The delta function with factor unity in the right-hand side of Eq.~\eqref{continuous_limit} suggests that ${\bf k}={\bf 0}$ is non-degenerate. 
The separation of the state ${\bf k}={\bf 0}$ is a standard step in the description of continuous systems~\cite{kubo1983statistical, landau2013course, kavokin2017microcavities}. 
Below we consider the quadratic dispersion, $\omega_{\bf k} = \omega_{\bf 0} + \alpha {\bf k}^2$.
This dispersion is characteristic for polaritons of the lower branch in the vicinity of~${\bf k} = {\bf 0}$ and the most frequent case in the experimental realizations of BEC.

The definition of the partition function~\eqref{partition_function_def} cannot be used directly to calculate $Z_N$ in the continuous limit.
The fastest way to calculate $Z_N$ is to use a recurrence relation~\cite{fraser1951xv, borrmann1993recursion, brosens1997thermodynamics, weiss1997particle, chase1999canonical, kocharovsky2006fluctuations}. 
The recurrence relation follows from the fact that the density matrix~\eqref{thermolized_dm_N} corresponds to exactly $N$ polaritons distributed in the system as a whole.
Thus,
\begin{equation}
	N = \sum_{\bf k} \langle \hat n_{\bf k} \rangle_N,
\end{equation}
where $\langle \hat n_{\bf k} \rangle_N$ stands for ${\rm Tr}( \hat n_{\bf k} \hat \rho _N)$.
According to Eqs.~\eqref{average} and~\eqref{derivative_Z_N},
\begin{equation}
	\sum_{\bf k} \langle \hat n_{\bf k} \rangle_N = 
	\sum_{n=1}^N \frac{Z_{N-n}}{Z_N} \sum_{\bf k}e^{-n(\omega_{\bf k}-\omega_{\bf 0})/{T}}.
\end{equation}
After integration over $\bf k$ we obtain the recurrence relations
\begin{equation} \label{recurrent_2D}
	Z_N = \frac{1}{N} \sum_{n=0}^{N-1} Z_n \left( 1 + \frac{G_{2D}}{N-n} \right)
\end{equation}
with $G_{2D} = VT/4\pi\alpha$ in 2D and
\begin{equation} \label{recurrent_3D}
	Z_N = \frac{1}{N} \sum_{n=0}^{N-1} Z_n \left( 1 + \frac{G_{3D}}{(N-n)^{3/2}} \right)
\end{equation}
with $G_{3D} = V(T/4\pi\alpha)^{3/2}$ in 3D.
In both cases, the initial value is $Z_0=1$ (see Eq.~\eqref{partition_function_0}).
The physical meaning of $G_{2D}$ and $G_{3D}$ is the number of states in the energy range $T$ above ${\bf k}={\bf 0}$ in the corresponding dimensions.

The exact expressions for $Z_N$ are obtained in Appendix~\ref{sec:appendix_a} trough Darwin--Fowler method~\cite{fraser1951xiv, fraser1951xv,  schubert1946bose, schubert1947bose, holthaus1999saddle}.
However, for practical calculations, the optimal strategy to calculate $Z_N$ is to use the recurrence relations~\eqref{recurrent_2D} and~\eqref{recurrent_3D}.

We can distinguish three regions in $\{ N, T \}$ plane and write the asymptotics for $Z_N$ in the corresponding regimes.

\paragraph{BEC is not formed.}
This regime is defined by $1 \ll N \ll G_{2D}$ in 2D and $1 \ll N \ll G_{3D}$ in 3D ($1 \ll N \ll T/T_0$ in the figures). 
From the Eqs.~\eqref{recurrent_2D}~and~\eqref{recurrent_3D} we obtain
\begin{equation} \label{Boltzman_ZN_2D}
Z_N \approx \frac{G_{2D}^N}{N!}
\end{equation}
in 2D and 
\begin{equation} \label{Boltzman_ZN_3D}
Z_N \approx \frac{G_{3D}^N}{N!}
\end{equation}
in 3D. 
These asymptotics corresponds to the ideal Boltzmann gas.
This is expected, since in this region the concentration of polaritons is small, therefore, the Bose effect does not have an influence on the distribution of the polaritons.

\paragraph{BEC is formed.}
This regime is defined by $1 \ll G_{2D} \ll N$ in 2D and $1 \ll G_{3D} \ll N$ in 3D ($1 \ll T/T_0 \ll N$ in the figures). 
From the Eqs.~\eqref{asymptotics_sum_2D}~and~\eqref{asymptotics_sum_3D} we obtain
\begin{multline} \label{Condensate_ZN_2D}
\sum_{n=0}^{N-1} Z_{n} \approx Z_\infty(N - G_{2D} \ln N) \;\Rightarrow\; \\ \;\Rightarrow\;  Z_N \approx Z_\infty \left( 1 - \frac{G_{2D}}{N} \right)
\end{multline}
with $Z_\infty=\exp(G_{2D}\zeta(2))$ in 2D and 
\begin{multline} \label{Condensate_ZN_3D}
\sum_{n=0}^{N-1} Z_{n} \approx Z_\infty(N - G_{3D} \zeta(3/2)) \;\Rightarrow\; \\ \;\Rightarrow\;  Z_N \approx Z_\infty \left( 1 - \frac{G_{3D}}{N^{3/2}} \right)
\end{multline}
with $Z_\infty=\exp(G_{3D}\zeta(5/2))$ in 3D, where $\zeta(x)$ is the zeta function evaluated at~$x$. 
This regime corresponds to the formation of BEC.

\paragraph{Deep quantum regime.}
This regime is defined by $G_{2D} \ll 1$ in 2D and $G_{3D} \ll 1$ ($T/T_0 \ll 1$ in the figures), when almost only ground state is populated.
From the Eqs.~\eqref{recurrent_2D}~and~\eqref{recurrent_3D} both in 2D and in 3D we obtain
\begin{equation} \label{Deep_quantum_ZN}
Z_N \approx 1.
\end{equation}

\section{Condensation of exactly $N$ polaritons} \label{sec:condensation}

\begin{figure*} [t] 
    \includegraphics[width=0.95\linewidth]{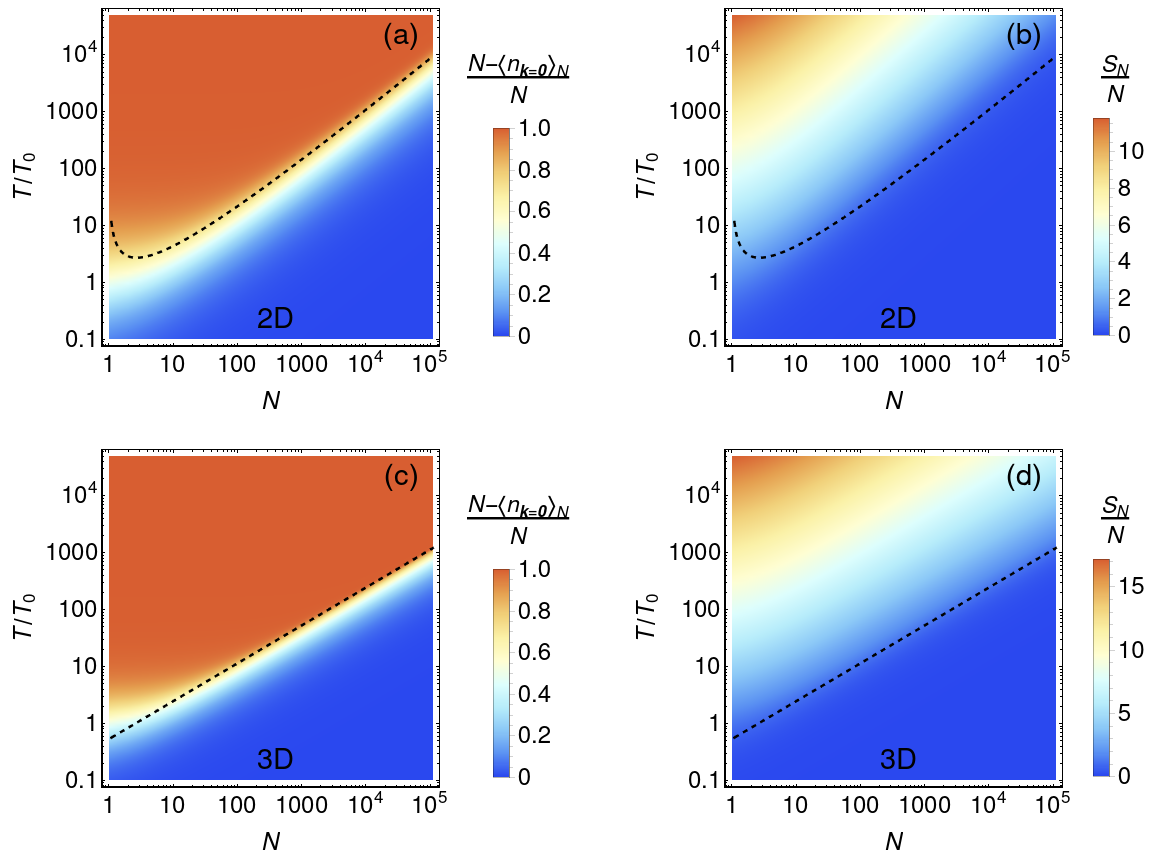}
    \caption{
The fraction of the polaritons in the excited states in 2D (a) and in 3D (c), given that there are exactly $N$ polaritons distributed in the system as a whole.
The entropy per polariton in 2D (b) and in 3D (d), given that there are exactly $N$ polaritons in the system.
$T$ is the temperature of the environment, $T_0$ is the temperature corresponding to $G_{2D}=1$ in 2D and $G_{3D}=1$ in 3D (see Section~\ref{sec:partition_function}).
The black dashed line marks the approximate condition for a macroscopic occupation of the ground state $N = G_{2D} \ln N$ in 2D (see Eq.~\eqref{2D_N=infty}) and $N = G_{3D} \zeta (3/2)$ in 3D (see Eq.~\eqref{3D_N=infty}).
}
    \label{fig:N_SN}
\end{figure*}

In Section~\ref{sec:density_matrix} we showed that in the fast thermalization limit the invariant subspaces play a substantial role in the dynamics of BEC.
In this section, we explore the formation of BEC in these invariant subspaces.
This means that in Eq.~\eqref{thermolized_dm} only one probability $P_N$ is non-zero.
Therefore, the system is described by the density matrix~\eqref{thermolized_dm_N} and has exactly $N$ polaritons distributed in the system as a whole.

\subsection{Population of the states}
From Eq.~\eqref{average} it follows that the average number of polaritons in the state with the wave vector $\bf k$ can be calculated as
\begin{equation}\label{n_k_N}
\langle \hat n_{\bf k} \rangle_N = \sum_{n=1}^N \frac{Z_{N-n}}{Z_N}e^{-n(\omega_{\bf k}-\omega_{\bf 0})/{T}}   .
\end{equation}
The subscript $N$ refers to the total number of polaritons distributed in the system as a whole.
This expression is the same both in 2D and in 3D. 
From the Fig.~\ref{fig:N_SN} one can see that the macroscopic occupation occurs for the fixed volume both in 2D and in 3D at sufficiently low temperatures. 

In the limit $N\to\infty$, from Eq.~\eqref{n_k_N} we obtain the fraction of polaritons in the ground state (see~Appendix~\ref{sec:appendix_b}) 
\begin{equation} \label{2D_N=infty}
\frac{\langle \hat n_{{\bf k}={\bf 0}} \rangle_N}{N} \approx
1 - \left(\frac{T}{4\pi\alpha}\right)\left(\frac{V}{N}\right)\ln(N)
\end{equation}
in 2D and 
\begin{equation} \label{3D_N=infty}
\frac{\langle \hat n_{{\bf k}={\bf 0}} \rangle_N}{N} \approx
1 - \left(\frac{T}{4\pi\alpha}\right)^{3/2}\left(\frac{V}{N}\right)\zeta(3/2)
\end{equation}
in 3D.

In the limit $N,V \to \infty $, $N/V \to \rm{const}$, the Eq.~\eqref{3D_N=infty} recovers the well-known textbook expression for BEC in 3D~\cite{kubo1983statistical, landau2013course, pitaevskii2003BEC}, while Eq.~\eqref{2D_N=infty} is in an agreement with the previously obtained results in 2D~\cite{ketterle1996bose, weill2019bose}.
In this limit, the fraction of the polaritons in the ground state becomes macroscopic in the 3D case as the $T$ decreases, however, in the 2D case, this fraction is infinitesimally small for any finite temperature.

\subsection{Entropy}
The density matrix~\eqref{thermolized_dm_N} allows obtaining the entropy, $S_N=-{\rm Tr}(\hat \rho_N\ln\hat\rho_N)$, in 2D
\begin{equation} \label{entropy_N_2D}
S_N = 
\ln Z_N + G_{2D}\sum_{n=1}^N\frac{Z_{N-n}}{Z_N}\frac{1}{n^2}
\end{equation}
and in 3D 
\begin{equation} \label{entropy_N_3D}
S_N = 
\ln Z_N + \frac{3}{2}G_{3D}\sum_{n=1}^N\frac{Z_{N-n}}{Z_N}\frac{1}{n^{5/2}}.
\end{equation}
The entropy per polariton decreases as the total number of polaritons grows~(Fig.~\ref{fig:N_SN}).
This is because an increase in the number of polaritons leads to an increase in the ratio of polaritons that are in the ground state.

\begin{figure*} [t]
    \includegraphics[width=\linewidth]{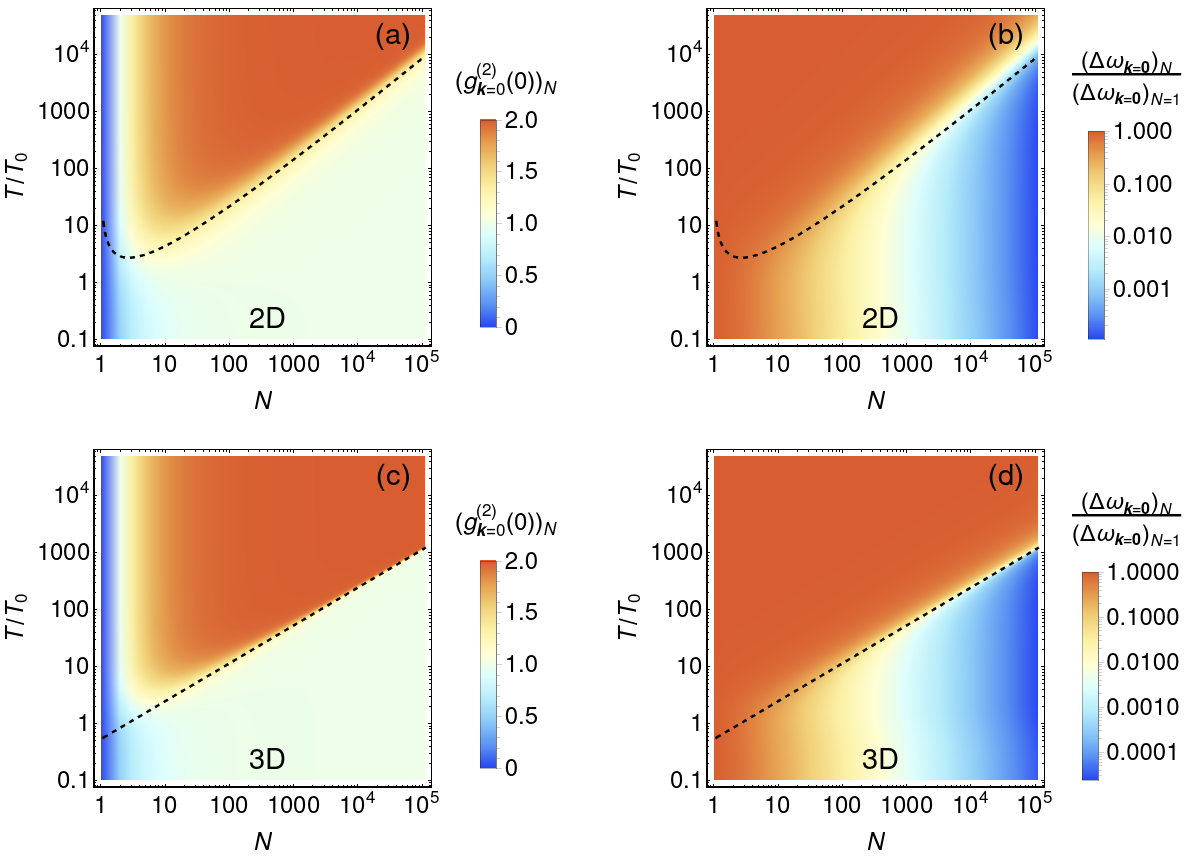}
    \caption{
The second-order coherence function for the polaritons in the ground state in 2D (a) and in 3D (c), given that there are exactly $N$ polaritons distributed in the system as a whole.
The spectral width of the emission from the polaritons in the ground state in 2D (b) and in 3D (d), given that there are exactly $N$ polaritons in the system.
$T$ is the temperature of the environment, $T_0$ is the temperature corresponding to $G_{2D}=1$ in 2D and $G_{3D}=1$ in 3D (see Section~\ref{sec:partition_function}).
The black dashed line marks the approximate condition for a macroscopic occupation of the ground state $N = G_{2D} \ln N$ in 2D (see Eq.~\eqref{2D_N=infty}) and $N = G_{3D} \zeta (3/2)$ in 3D (see Eq.~\eqref{3D_N=infty}).
}
    \label{fig:g2_SpectralWidth}
\end{figure*}

\subsection{First-order coherence function}
The first-order coherence function is defined as~\cite{doan2008coherence, scully1999quantum}
\begin{equation} \label{g1_N}
g^{(1)}_N({\bf r}) = 
\frac
{\langle\hat\psi^\dag({\bf 0})\hat\psi({\bf r})\rangle_N}
{\sqrt{\langle\hat\psi^\dag({\bf 0})\hat\psi({\bf 0})\rangle}_N \sqrt{\langle\hat\psi^\dag({\bf r})\hat\psi({\bf r})\rangle}_N},
\end{equation}
where $\hat \psi({\bf r})$ is a wave function with the following plane-wave expansion:
\begin{equation}
\hat\psi({\bf r})=
	\frac{1}{\sqrt{V}}\sum_{\bf k} \hat a_{\bf k} e^{i {\bf k}{\bf r}}.
\end{equation}
From Eq.~\eqref{thermolized_dm_N} it follows that $\langle {\hat a_{\bf{k}}^\dag {{\hat a}_{{\bf{k'}}}}} \rangle_N  = \left\langle {{{\hat n}_{\bf{k}}}} \right\rangle_N {\delta _{{\bf{kk'}}}}$.
Thus, we obtain
\begin{multline} \label{g1_N_2D}
g^{(1)}_N({\bf r}) = 
	\frac{\langle \hat n_{{\bf k}={\bf 0}}  \rangle_N}{N} + \\+
	\frac{G_{2D}}{N} \sum_{n=1}^N\frac{Z_{N-n}}{Z_N}\frac{1}{n} e^{ -r^2T/4\alpha n }
\end{multline}
in 2D and 
\begin{multline} \label{g1_N_3D}
g^{(1)}_N({\bf r}) = 
	\frac{\langle \hat n_{{\bf k}={\bf 0}}  \rangle_N}{N} + \\+
	\frac{G_{3D}}{N} \sum_{n=1}^N\frac{Z_{N-n}}{Z_N}\frac{1}{n^{3/2}} e^{ -r^2T/4\alpha n }
\end{multline}
in 3D.
The definition~\eqref{g1_N} implies that ${g^{(1)}_N({\bf 0})=1}$.

Above the condensation threshold, the term $\langle \hat n_{{\bf k}={\bf 0}} \rangle_N / N$ does not vanish, therefore, a non-equilibrium transition to BEC is always accompanied by the buildup of long-range spatial correlations.
We note, that this result does not contradict the Mermin--Wagner theorem, that was initially formulated for the magnetic systems~\cite{mermin1966absence}.
In 3D, the condensation threshold can be reached both for finite $V$ and infinite $V$, that results in the buildup of the long-range correlations.
However, in 2D, the buildup of long-range correlations is possible only for finite $V$, but not for infinite $V$.
Indeed, according to~\eqref{2D_N=infty} the term $\langle \hat n_{{\bf k}={\bf 0}}\rangle_N/N$ vanishes in the limit $N,V \to \infty $, $N/V \to \rm{const}$.
Thus, for infinite sized 2D Bose gas, the condensation threshold cannot be reached and the long-range correlations cannot arise.

Below the condensation threshold, $g^{(1)}_N({\bf r})=0$ as $r\to+\infty$.
This decay of second-order coherence function is exponential in the limit $N \to \infty$, $V \to \infty$, $N/V \to {\rm const}$ with the coherence length 
\begin{equation} \label{coherence_length_2D}
l_{2D}=\sqrt{\frac{\alpha}{T}\frac{1}{\ln \left[ (V/N)(T/4 \pi \alpha) \right]}}
\end{equation}
in 2D and 
\begin{equation} \label{coherence_length_3D}
l_{3D}=\sqrt{\frac{\alpha}{T}\frac{1}{\ln \left[ (V/N)(T/4 \pi \alpha)^{3/2} \right]}}
\end{equation}
in 3D.
Indeed, from Eqs.~\eqref{g1_N_2D}~and~\eqref{g1_N_3D} by using expressions~\eqref{Boltzman_ZN_2D}--\eqref{Boltzman_ZN_3D}, replacing the sum over $n$ by the integral and applying the Laplace's method, we obtain 
\begin{equation} \label{asymptotic_g1_2D}
g^{(1)}_N({\bf r}) 
\approx
\frac{G_{2D}}{N} \sqrt{\frac{2 \pi l_{2D}}{r}} e^{ -{r}/{l_{2D}} }
\end{equation}
for $r \gg l_{2D}$ in 2D and 
\begin{equation} \label{asymptotic_g1_3D}
g^{(1)}_N({\bf r}) 
\approx
\frac{G_{3D}}{N} \sqrt{4\pi \ln \left( \frac{G_{3D}}{N} \right)} \frac{l_{3D}}{r} e^{ -{r}/{l_{3D}} } 
\end{equation}
for $r \gg l_{3D}$ in 3D.
The asymptotic behaviors described by Eq.~\eqref{asymptotic_g1_2D}~and~\eqref{asymptotic_g1_3D} are in an agreement with~\cite{pitaevskii2003BEC, mouchliadis2008first}.

\subsection{Second-order coherence function}
Using Eq.~\eqref{average}, we obtain the second-order coherence function of the ground state
\begin{equation} \label{g2_ground}
(g^{(2)}_{{\bf k}={\bf 0}})_N=
	\frac
	{
	\langle \hat n_{{\bf k}={\bf 0}}\hat n_{{\bf k}={\bf 0}}\rangle_N - \langle \hat n_{{\bf k}={\bf 0}}\rangle_N
	}
	{
	\langle \hat n_{{\bf k}={\bf 0}}\rangle_N^2
	},
\end{equation}
where $ \langle\hat n_{{\bf k}={\bf 0}}\hat n_{{\bf k}={\bf 0}}\rangle_N = \sum_{n=1}^N Z_{N-n}(2n-1)/Z_N $. 
Eqs.~\eqref{g2_ground}~and~\eqref{Deep_quantum_ZN}--\eqref{Condensate_ZN_3D} allow us to consider the second-order coherence function in some limiting cases.
In the deep quantum regime ($T/T_0 \ll 1$ in Fig.~\ref{fig:g2_SpectralWidth}), $(g^{(2)}_{{\bf k}={\bf 0}}(0))_N \approx 1-1/N$, that corresponds to a Fock state.
When BEC is not formed ($1 \ll N \ll T/T_0$  in Fig.~\ref{fig:g2_SpectralWidth}), $(g^{(2)}_{{\bf k}={\bf 0}}(0))_N \approx 2$, that corresponds to an incoherent state.
When BEC is formed ($1 \ll T/T_0 \ll N$  in Fig.~\ref{fig:g2_SpectralWidth}), $(g^{(2)}_{{\bf k}={\bf 0}}(0))_N \approx 1 + G_{2D}/N$ in 2D and $(g^{(2)}_{{\bf k}={\bf 0}}(0))_N \approx 1 + 4 G_{3D}/N^{3/2}$ in 3D, that corresponds to a system state tends to coherent state.

Thus, we conclude, that for a fixed total number of polaritons, the macroscopic occupation of the ground state is accompanied by a buildup of coherence~(see Fig.~\ref{fig:N_SN}~and~Fig.~\ref{fig:g2_SpectralWidth}).
When $P_N$ (see Eq.~\eqref{thermolized_dm}) is well localized over some $N_0 \gg 1$, the analysis presented above for exactly $N_0$ polaritons distributed in the system as a whole is a good physical approximation.
However, in many cases, the distribution of $P_N$ and, consequently, pumping scheme significantly alters the coherent properties of the BEC.
This information can be accessed by the consideration of the kinetics of the condensation (see Section~\ref{sec:kinetics}).

\subsection{Spectral width} \label{sec:spectrum}
The stationary emission spectrum, $I_{\bf k}(\omega)$, of the polaritons with the wave vector $\bf k$ is determined by the correlation function $\langle \hat a_{\bf k}^\dag(t) \hat a_{\bf k}(t+\tau)\rangle$~\cite{scully1999quantum, breuer2002theory} 
\begin{equation} \label{spectrum}
I_{\bf k}(\omega)=
\lim_{t\to +\infty}{\rm Re} \int_0^{+\infty} \langle \hat a_{\bf k}^\dag(t) \hat a_{\bf k}(t+\tau)\rangle e^{-i\omega \tau} d\tau.
\end{equation}
According to the quantum regression theorem~\cite{scully1999quantum, breuer2002theory}, $\langle \hat a_{\bf k}^\dag(t) \hat a_{\bf k}(t+\tau)\rangle = {\rm Tr} \left( \hat a_{\bf k} \hat\rho(\tau) \hat a_{\bf k}^\dag\right)$ at $t \to \infty$ and 
\begin{multline} \label{Quantum_regression}
\frac{d}{d\tau}{\rm Tr} \left( \hat a_{\bf k} \hat\rho(\tau) \hat a_{\bf k}^\dag\right) 
= {\rm Tr}\left( \hat a_{\bf k} L_{\rm pump}(\hat \rho(\tau) \hat a_{\bf k}^\dag) \right)  \\
+ {\rm Tr}\left( \hat a_{\bf k} L_{\rm diss}(\hat \rho(\tau) \hat a_{\bf k}^\dag) \right) 
+ {\rm Tr}\left( \hat a_{\bf k} L_{\rm therm}(\hat \rho(\tau) \hat a_{\bf k}^\dag) \right), 
\end{multline}
where at the initial moment in time $\hat \rho(0)$ is defined by the stationary solution~\eqref{solution_P_1}--\eqref{solution_P_N}.
To obtain $\langle \hat a_{\bf k}^\dag(t) \hat a_{\bf k}(t+\tau)\rangle$, we solve the Eq.~\eqref{Quantum_regression} in the matrix form and then find the average $\langle \hat a_{\bf k}^\dag(t) \hat a_{\bf k}(t+\tau)\rangle$. 
In the fast thermalization limit, we may neglect $\hat a_{\bf k} L_{\rm pump}(\hat \rho(\tau) \hat a_{\bf k}^\dag)$ and $\hat a_{\bf k} L_{\rm diss}(\hat \rho(\tau) \hat a_{\bf k}^\dag)$ because these terms are small compared to $\hat a_{\bf k} L_{\rm therm}(\hat \rho(\tau) \hat a_{\bf k}^\dag)$.
As a result, we obtain the following equation 
\begin{equation} \label{Quantum_regression_matrix_approx}
\hat a_{\bf k} \frac{d\hat\rho(\tau) }{d\tau} \hat a_{\bf k}^\dag
\approx \hat a_{\bf k} L_{\rm therm}(\hat \rho(\tau)) \hat a_{\bf k}^\dag +
L_{\rm therm}'(\hat \rho(\tau)) , 
\end{equation}
where
\begin{multline} 
L_{\rm therm}'(\hat \rho) =
\hat a_{\bf k}\hat a_{\bf k}^\dag \sum_{{\bf q}\neq{\bf k}} \Gamma_{\bf qk} \hat a_{\bf q}\hat \rho\hat a_{\bf q}^\dag \\ +
\frac{1}{2} \sum_{{\bf q}\neq{\bf k}} \Gamma_{\bf kq} \hat a_{\bf k} \hat \rho \hat a_{\bf q} \hat a_{\bf q}^\dag \hat a_{\bf k}^\dag +
\frac{1}{2} \sum_{{\bf q}\neq{\bf k}} \Gamma_{\bf qk} \hat a_{\bf k} \hat \rho \hat a_{\bf q}^\dag \hat a_{\bf q} \hat a_{\bf k}^\dag .
\end{multline}
To solve this equation, we assume that $L_{\rm therm}$ in the right-hand side of Eq.~\eqref{Quantum_regression_matrix_approx} leads to a decay with the characteristic thermalization rate $\Gamma$.
We also assume that $L'_{\rm therm}$ leads to a decay with a rate much lower than $\Gamma$, but much higher than the dissipation rate and the pumping rate. 
After obtaining the solution of Eq.~\eqref{Quantum_regression_matrix_approx}, we explicitly check these assumptions.
Employing these assumptions, we search for a matrix $\rho(\tau)$ satisfying Eq.~\eqref{Quantum_regression_matrix_approx} in the form~\eqref{thermolized_dm}.
As a result, we obtain 
\begin{equation} \label{correlators_for_spectrum}
	\langle \hat a_{\bf k}^\dag(t) \hat a_{\bf k}(t+\tau)\rangle_N = \langle \hat n_{\bf k}(t)\rangle_N e^{-i\omega_{\bf k} \tau-(\Delta\omega_{\bf k})_N\tau/2}
\end{equation}
as $t\to+\infty$, where 
\begin{multline} \label{Gamma_N}
(\Delta\omega_{\bf k})_N
 = 
\sum_{{\bf q}\neq{\bf k}} \Gamma_{\bf kq} - \\ -
\sum_{{\bf q}\neq{\bf k}} (\Gamma_{\bf qk} - \Gamma_{\bf kq}) \frac{\langle (1+\hat n_{\bf k}) \hat n_{\bf q} \rangle_{N-1}}{\langle 1 + \hat n_{\bf k} \rangle_{N-1}}.
\end{multline}

We consider in more detail the linewidth of the ground state, $(\Delta\omega_{{\bf k}={\bf 0}})_N$.
For simplicity, we assume that the thermalization rates do not depend on the states, i.e., $\Gamma_{\bf q0}=\Gamma$.
Note, that the thermalization rates obey the Kubo--Martin--Schwinger relation, $\Gamma_{\bf 0q}=\Gamma\exp(-(\omega_{\bf q}-\omega_{\bf 0})/T)$.
From Eq.~\eqref{average}, \eqref{density_of_states} and~\eqref{Gamma_N} we obtain 
\begin{equation} \label{Gamma_N_omega}
	(\Delta\omega_{{\bf k}={\bf 0}})_{N} = 
\Gamma 
\frac{NZ_N-\sum_{n=0}^{N-1} Z_{n} }{\sum_{n=0}^{N-1} Z_{n}},
\end{equation}
When $T$ is higher than the condensation temperature, the linewidth of the ground state emission is almost independent of the total number of polaritons~(Fig.~\ref{fig:g2_SpectralWidth}). 
But, when $T$ is lower than the condensation temperature, an increase in $N$ leads to the narrowing of the linewidth~(Fig.~\ref{fig:g2_SpectralWidth}).
In the letter case ($N \gg G_{2D}$ in 2D and $N \gg G_{3D}$ in 3D), we use Eqs.~\eqref{Deep_quantum_ZN},~\eqref{Condensate_ZN_2D}~and~\eqref{Condensate_ZN_3D} to obtain
\begin{equation} \label{Gamma_N_omega_2D_lowT}
(\Delta\omega_{{\bf k}={\bf 0}})_{N} \approx 
\Gamma G_{2D} \frac{\ln N}{N}
\end{equation}
in 2D and 
\begin{equation} \label{Gamma_N_omega_3D_lowT}
(\Delta\omega_{{\bf k}={\bf 0}})_{N} \approx 
\Gamma G_{3D} \frac{\zeta(3/2)}{N}
\end{equation}
in 3D.
Thus, in 2D, the linewidth of the ground state deviates from the Schawlow--Townes law, whereas, in 3D, the linewidth follows the Schawlow--Townes law.
The influence of the dimensionality on the Schawlow--Townes narrowing is unexpected effect.
As far as we know, this effect reported here for the first time.
It is interesting to note, that the linewidth of the emitted light for the ground state~\eqref{Gamma_N_omega_2D_lowT} and~\eqref{Gamma_N_omega_3D_lowT} is proportional to the fraction of the polaritons in the excited states (see Eqs.~\eqref{2D_N=infty}~and~~\eqref{3D_N=infty}).

From Eqs.~\eqref{spectrum}~and~\eqref{correlators_for_spectrum} it follows that $\langle \hat a_{\bf k}^\dag(t) \hat a_{\bf k}(t+\tau)\rangle = \sum_N P_N \langle \hat a_{\bf k}^\dag(t) \hat a_{\bf k}(t+\tau)\rangle_N$, therefore, the linewidth depends on the distribution of the probabilities $P_N$.
Eq.~\eqref{Gamma_N} is a good physical approximation for a linewidth if the distribution $P_N$ has a sharp peak over some $N_0 \gg 1$.
In this case, the linewidth of the emission is Lorentzian.
In other cases, the kinetics of polaritons may alter the shape of the spectrum of the emitted light.

\section{Kinetics of the condensation} \label{sec:kinetics}
Generally, there are two conditions for the formation of the condensate.
The first condition is the macroscopic occupation of the ground state.
This condition means that $\langle \hat N \rangle$ and $T$ are in the condensation region in the $\{ N, T\}$ plane~(see Fig.~\ref{fig:N_SN}). 
The second condition is the buildup of coherence in the ground state.
This condition means that the second-order coherence function reaches $1$. 
Below, we show that these two conditions differ in non-equilibrium BEC in the fast thermalization limit.

Due to dissipation and pumping processes, all the probabilities~$P_N$ differ from zero and may depend on time. 
To obtain~$P_N$ we substitute the thermalized density matrix~\eqref{thermolized_dm} into Eq.~\eqref{Master_equation}.
As a result, we obtain
\begin{equation} \label{equation_P_0}
\frac{\partial P_0(t)}{\partial t} = \frac{d_0}{Z_1}P_1(t) - \frac{p_0}{Z_0}P_0(t),
\end{equation}
\begin{multline} \label{equation_P_N}
\frac{\partial P_N(t)}{\partial t} = \frac{d_N}{Z_{N+1}}P_{N+1}(t) -\\ -
	\frac{d_{N-1}+p_N}{Z_N}P_N(t) + \frac{p_{N-1}}{Z_{N-1}}P_{N-1}(t)
\end{multline}
for $N>0$. 
The coefficients $p_N$ and $d_N$ are defined by 
\begin{equation}
	{d_N} = Z_{N+1} \sum_{\bf k} \left( {{\gamma _{\bf{k}}} + {\kappa _{\bf{k}}}} \right) \langle \hat n_{\bf k} \rangle_{N+1},
\end{equation}
\begin{equation}
	{p_N} = Z_{N+1} \sum_{\bf k} {\kappa _{\bf{k}}}  e^{(\omega_{\bf k} - \omega_{\bf 0})/T}  \langle \hat n_{\bf k} \rangle_{N+1}.
\end{equation}

If dissipation and pumping rates do not change in time, then the stationary solution of Eqs.~\eqref{equation_P_0}--\eqref{equation_P_N} can be obtained in the form of the recurrence relation
\begin{equation} \label{solution_P_1}
P_1 = \frac{Z_1}{Z_0}\frac{p_0}{d_0}P_0,
\end{equation} 
\begin{equation} \label{solution_P_N}
P_{N+1} = \frac{Z_{N+1}}{Z_N}\frac{d_{N-1}+p_N}{d_N}P_N - \frac{Z_{N+1}}{Z_{N-1}}\frac{p_{N-1}}{d_N}P_{N-1}
\end{equation}
for $N\ge 1$.
The probability $P_0$ in~\eqref{solution_P_1} can be obtained from the normalization condition $\sum_{N=0}^{+\infty}P_N=1$.

Once we know the probabilities $P_N$, we can find the reduced density matrix, $\hat \rho_{\bf k}$, of any state with the wave vector $\bf k$. 
This can be done by tracing out all the states in the Eq.~\eqref{thermolized_dm} except for $\bf k$ (see Appendix~\ref{sec:appendix_c})
\begin{multline} \label{reduced_dm}
\hat \rho_{\bf k} = \sum_{n=0}^{+\infty} |n\rangle \langle n| e^{-n(\omega_{\bf k}-\omega_{\bf 0})/T} \times
\\
\times
\sum_{N=n}^{+\infty} P_N\frac{Z_{N-n} - Z_{N-n-1}e^{-(\omega_{\bf k}-\omega_{\bf 0})/T}}{Z_N}  
\end{multline}
where state $| n \rangle$ satisfies $\hat n_{\bf k} | n \rangle = n | n \rangle$ and we formally put $Z_{-1}=0$.
The Eq.~\eqref{reduced_dm} is in agreement with the previously obtained results~\cite{wilkens1997particle, weiss1997particle, mossel2012exact}.
The reduced density matrix of the ground state has a particularly simple form
\begin{equation} \label{reduced_dm_gs}
\hat \rho_{{\bf k}={\bf 0}} = \sum_{n=0}^{+\infty} |n\rangle \langle n| 
\sum_{N=n}^{+\infty} P_N\frac{Z_{N-n} - Z_{N-n-1}}{Z_N}  
\end{equation}
Since Eq.~\eqref{reduced_dm_gs} is valid for any set of probabilities $P_N$, then the positive semi-definiteness of $\hat \rho_{{\bf k}={\bf 0}}$ requires $Z_{N+1} \geq Z_N$ for any $N$. 
In Appendix~\ref{sec:appendix_a}, we proof that $Z_N$ is monotonic.
We emphasize, that monotonicity of $Z_N$ is a direct consequence of the ground state separation, that we did in the Eq.~\eqref{continuous_limit} (see also Eqs.~\eqref{generating_function},~\eqref{Z_N_and_F_N}~and~\eqref{F_N}).

From Eq.~\eqref{thermolized_dm} it follows, that both the particle number in the ground state $\langle \hat n_{{\bf k}={\bf 0}} \rangle$ and the entropy $S$ can be expressed through the probabilities $P_N$:
\begin{equation} \label{n_0}
\langle \hat n_{{\bf k}={\bf 0}} \rangle=\sum_{N=0}^{+\infty} P_N \langle \hat n_{{\bf k}={\bf 0}} \rangle_N,
\end{equation} 
\begin{equation} \label{entropy}
S=\sum_{N=0}^{+\infty} P_N S_N - \sum_{N=0}^{+\infty} P_N \ln P_N,
\end{equation} 
where $\langle \hat n_{{\bf k}={\bf 0}} \rangle_N$ is defined by Eq.~\eqref{n_k_N} and $S_N$ is defined by Eq.~\eqref{entropy_N_2D} in 2D and by Eq.~\eqref{entropy_N_3D} in 3D. 
This means that the total average number of polaritons is the sum over the invariant subspaces weighted with the probabilities of the system being in the corresponding subspace.

Unlike the average number of polaritons in the ground state, $g^{(1)}({\bf r})$ and $g^{(2)}_{{\bf k}={\bf 0}}(0)$ are not the weighted sum of $g^{(1)}_N({\bf r})$ and $( g^{(2)}_{{\bf k}={\bf 0}}(0) )_N$ over invariant subspaces, but instead
\begin{widetext}
\begin{equation} \label{g1}
g^{(1)}({\bf r}) = 
\frac
{\sum_{N=0}^{+\infty} P_N\langle\hat\psi^\dag({\bf 0})\hat\psi({\bf r})\rangle_N}
{\sqrt{\sum_{N=0}^{+\infty} P_N\langle\hat\psi^\dag({\bf 0})\hat\psi({\bf 0})\rangle_N} \sqrt{\sum_{N=0}^{+\infty} P_N\langle\hat\psi^\dag({\bf r})\hat\psi({\bf r})\rangle_N}},
\end{equation}
\end{widetext}
\begin{equation} \label{g2_0}
g^{(2)}_{{\bf k}={\bf 0}}(0) = 
\frac{\sum_{N=0}^{+\infty} P_N \langle \hat n_{{\bf k}={\bf 0}} \hat n_{{\bf k}={\bf 0}} - \hat n_{{\bf k}={\bf 0}} \rangle_N}{\left( \sum_{N=0}^{\infty} P_N \langle \hat n_{{\bf k}={\bf 0}} \rangle_N \right)^2}.
\end{equation}
Therefore, the coherence of the condensate is determined not only by $\langle \hat N \rangle$ and $T$, but also strongly depends on the particular distribution of the $P_N$. 

Below we consider in more detail the buildup of the coherence of stationary non-equilibrium BEC at $T=0$ in the fast thermalization limit.

\paragraph{Coherence buildup at $T=0$.} 
As it was shown in Section~\ref{sec:condensation}, $\langle \hat n_{{\bf k}={\bf 0}} \rangle_N=N$, $\langle \hat n_{{\bf k}\neq{\bf 0}} \rangle_N=0$ and $(g^{(2)}_{{\bf k}={\bf 0}})_N=1-N^{-1}$ at $T=0$ both in 2D and in 3D.
This means that the polaritons in ${\bf k} = {\bf 0}$ in each of the invariant subspace are in the Fock state.  
Nevertheless, this does not imply a buildup of the coherence because the dissipation and the pumping processes should be taken into account.

The coherence of the polaritons in the ground state is completely determined by the reduced density matrix of the ground state, $\hat \rho_{{\bf k}={\bf 0}}$.
From Eq.~\eqref{reduced_dm_gs}, it follows that this reduced density matrix at~${T=0}$ takes the form 
\begin{equation} \label{rho_0_T=0}
	\hat \rho_{{\bf k}={\bf 0}} = \sum_{n=0}^{+\infty}|n\rangle\langle n| P_n.
\end{equation}
The stationary solution~\eqref{solution_P_1}--\eqref{solution_P_N} for probabilities $P_N$, in this case, reduces to
\begin{multline} \label{P_N_T=0}
P_N=
\frac{1}{N!}
\left( \frac{\kappa_{{\bf k}={\bf 0}}}{\gamma_{{\bf k}={\bf 0}} + \kappa_{{\bf k}={\bf 0}}} \right)^N \\
\times \left( 1 - \frac{\kappa_{{\bf k}={\bf 0}}}{\gamma_{{\bf k}={\bf 0}}+\kappa_{{\bf k}={\bf 0}}} \right)^{-\left( 1 + \kappa_1/\kappa_{{\bf k}={\bf 0}} \right)}\\
\times f_N\left( 1 + \frac{\kappa_1}{\kappa_{{\bf k}={\bf 0}}} \right),
\end{multline}
where $\gamma_{{\bf k}={\bf 0}}$ and $\kappa_{{\bf k}={\bf 0}}$ are the dissipation and the pumping rates for ${\bf k}={\bf 0}$, and $\kappa_1$ is the total pumping rate of the all excited states
\begin{equation} \label{kappa_1}
\kappa_1 =
\sum_{{\bf k}\neq{\bf 0}} \kappa_{\bf k}.
\end{equation}
The function $f_N$ is determined by
\begin{equation} \label{f_0}
f_0(x) = 1,
\end{equation}
\begin{equation} \label{f_N}
f_N(x) = x (x + 1) ... (x + N - 1)
\end{equation}
for $N>0$.

The reduced density matrix~\eqref{rho_0_T=0} and probabilities~\eqref{P_N_T=0} allow us to find the occupation of the ground state, first- and second-order correlation functions of the polaritons
\begin{equation} \label{n_0_T=0}
\langle \hat n_{{\bf k}={\bf 0}} \rangle =
\frac{\kappa_{{\bf k}={\bf 0}} + \kappa_1}{\gamma_{{\bf k}={\bf 0}}},
\end{equation}
\begin{equation} \label{g1_T=0}
g^{(1)}({\bf r}) = 1,
\end{equation}
\begin{equation} \label{g2_0_T=0}
g^{(2)}_{{\bf k}={\bf 0}}(0) = 
\frac{2\kappa_{{\bf k}={\bf 0}} + \kappa_1}{\kappa_{{\bf k}={\bf 0}}+\kappa_1}.
\end{equation}
This means that regardless of which polariton state is pumped, all the polaritons are thermalized to the ground state and long-range spatial correlations are established.
But different pumping schemes result in different statistics of the condensate.
Depending on the pumping scheme, the second-order coherence function of the ground state may be between $1$ and $2$.

When only the ground state is pumped incoherently, the coherence buildup does not occur at $T=0$.
This is because the transitions from the ground state to the excited states are suppressed and the ground state is effectively decoupled from the excited states at $T=0$.
As a result, the ground state inherits the statistics of the incoherent pumping.

The more the excited states are pumped compared to the ground state the more the polariton condensate is coherent.
In the limiting case $\kappa_{{\bf k}={\bf 0}}=0$, $\kappa_1\neq0$, the polaritons are perfectly coherent at $T=0$ and the reduced density matrix~\eqref{rho_0_T=0} becomes
\begin{equation} \label{rho_0_T=0_kappa_0=0}
\hat \rho_{{\bf k}={\bf 0}} = \sum_{n=0}^{+\infty}|n\rangle\langle n| 
\frac{1}{n!}
\left( \frac{\kappa_1}{\gamma_{{\bf k}={\bf 0}}} \right)^n
e^{-\kappa_1/\gamma_{{\bf k}={\bf 0}}}.
\end{equation} 
This state is a randomly phased coherent state.
The thresholdless condensation into a perfect coherent state at $T=0$, in this case, is a consequence of the fast thermalization limit.
Indeed, the critical number of particles necessary for the coherence buildup is determined by the ratio between thermalization rate and the dissipation rate~\cite{laussy2004spontaneousPSSC}. 
We consider the fast thermalization limit, therefore, this critical number of particles is almost zero.
Thus, we have a thresholdless transition to a perfect coherent state~\eqref{rho_0_T=0_kappa_0=0}.

\section{Conclusion} \label{sec:conclusion}
In conclusion, we developed an analytical framework to describe non-equilibrium ideal Bose--Einstein condensates (BEC) in the fast thermalization limit.
In this limit, we obtained an expression for the full density matrix of non-equilibrium ideal BEC.

We showed that a macroscopic occupation of the ground state occurs under the same conditions for the full thermal equilibrium case as for the non-equilibrium case in the fast thermalization limit.
The macroscopic occupation of the ground state is always accompanied by the formation of long-range first-order spatial correlations.
Moreover, for a given system, only the temperature and the average number of the particles determine whether the ground state is macroscopically occupied. 
In contrast, the buildup of second-order coherence in BEC is not fully determined by the occupation of the ground state, but also depends on the pumping scheme.
For instance, at zero temperature, if excited states are pumped, then a coherence of the ground state is formed, but if only the ground state is pumped, then the coherence buildup of ground state does not occur.

At fixed temperature, an increase in the number of polaritons leads to an increase in the total entropy, but the entropy per polariton decreases.
This is because the polaritons in the ground state have low entropy and an increase in the number of particles leads to an increase of the fraction of the particles in the ground state.

Above the condensation threshold, the linewidth of the ground state emission narrows.
At low temperatures and large numbers of polaritons, this narrowing follows the Schawlow--Townes law in 3D, but, in 2D, the linewidth decreases slower than predicted by the Schawlow--Townes law.
As the temperature decreases, the condensate line width decreases faster in 3D than in 2D. 

In this work, we considered only 2D and 3D polaritons with quadratic dispersion and non-degenerate states enumerated by the wave vectors.
However, the developed theory can be straightforwardly adopted to other dispersions, such as dispersion, with a bottleneck, an inflection point and degenerate states with definite wave vectors. 
Moreover, the description of the kinetics is tolerant to the more realistic situation, than was considered in this work, for instance, when radiative lifetime is bound by Hopfield expressions.   

The developed theory is a theory for the ideal gas.
However, this theory can be applied to non-ideal gases, with the Hamiltonian $\hat H + \hat U$, where $\hat U$ is non-ideal contribution to the energy.
If the operator $\hat U$ commutes with the operator of the total number of polaritons $\sum_{\bf k} \hat n_{\bf k}$, then the developed theory can be directly applied by replacing the partition function for the ideal gas $Z_N$ with $Z_N \exp(-\langle \hat U \rangle_N / T)$.
However, if the non-ideal part of the energy operator does not commute with the total number of polaritons, then the generalization of the presented theory might be limited to the first-order corrections.

\section*{Acknowledgments} \label{sec:Acknowledgements}
This work was supported by the Russian Science Foundation (Grant No. 20-72-10145). 
E.S.A. and V.Yu.Sh. thank the Foundation for the Advancement of Theoretical Physics and Mathematics ``Basis''.
Yu.E.L. acknowledges the Basic Research Program at the National Research University HSE.

\bibliographystyle{unsrtnat}
\bibliography{ThermalizationContinuous}

\onecolumn\newpage
\appendix

\section{Some properties of the partition function $Z_N$} \label{sec:appendix_a}
In this Appendix, we present some general properties of the partition function $Z_N$.
To derive these properties, we introduce the generating function of $Z_N$: $\sum_{N=0}^{+\infty}x^NZ_N$, $0 < x < 1$.
Eq.~\eqref{partition_function_def} allows us to represent the generating function in different forms 
\begin{multline} \label{generating_function}
\sum_{N=0}^{+\infty}x^NZ_N 
=
\sum_{N=0}^{+\infty}
x^N
\sum_{ {\rm config} : \sum_{\bf k} n_{\bf k}=N } 
e^{- \sum_{\bf k} n_{\bf k}(\omega_{\bf k}-\omega_{\bf 0})/{T}}
=
\sum_{N=0}^{+\infty}
\sum_{ {\rm config} : \sum_{\bf k} n_{\bf k}=N } 
e^{- \sum_{\bf k} n_{\bf k}((\omega_{\bf k}-\omega_{\bf 0})/{T} - \ln x)}
=\\=
\prod_{\bf k}
\sum_{n_{\bf k}=0}^{+\infty}
e^{- n_{\bf k}((\omega_{\bf k}-\omega_{\bf 0})/{T} - \ln x)}
=
\prod_{\bf k}
\frac{1}{1 - xe^{- (\omega_{\bf k}-\omega_{\bf 0})/{T}}}
=
\exp \left\{  
- \sum_{\bf k} \ln \left[ 1 - x e^{-(\omega_{\bf k}-\omega_{\bf 0})/T} \right]
\right\}.
\end{multline}
The last two forms are especially useful for the considerations below.

\subsection{Proof of Eq.~\eqref{average}}
From Eq.~\eqref{generating_function} it follows that
\begin{multline}
	\sum_{N=0}^{+\infty}x^N \frac{\partial Z_N}{\partial \omega_{\bf q}} 
	=
	-\frac{1}{T} \frac{x e^{-(\omega_{\bf q}-\omega_{\bf 0})/T}}{1 - x e^{-(\omega_{\bf q}-\omega_{\bf 0})/T}} 
	\prod_{\bf k} \frac{1}{1 - xe^{- (\omega_{\bf k}-\omega_{\bf 0})/{T}}}
	=\\=
	-\frac{1}{T} \frac{x e^{-(\omega_{\bf q}-\omega_{\bf 0})/T}}{1 - x e^{-(\omega_{\bf q}-\omega_{\bf 0})/T}} 
	\sum_{N=0}^{+\infty}x^NZ_N
	=
	\sum_{N=0}^{+\infty}x^N \frac{1}{T} \left( Z_N - \sum_{n=0}^N Z_{N-n}e^{n(\omega_{\bf q}-\omega_{\bf 0})/T} \right).
\end{multline}
Eq.~\eqref{average} directly follows from this equation.

\subsection{Monotonicity of $Z_N$}
To prove the monotonicity of $Z_N$ we use Eq.~\eqref{generating_function} and consider the expression
\begin{equation} \label{Z_N_and_F_N}
	\sum_{N=0}^{+\infty}x^NZ_N 
	=
	\frac{1}{1-x}\sum^{+\infty}_{N=0} x^N F_N,
\end{equation}
where we introduced
\begin{equation} \label{F_N}
	\sum^{+\infty}_{N=0} x^N F_N
	=
	\prod_{{\bf k}\neq{\bf 0}}
	\frac{1}{1 - xe^{- (\omega_{\bf k}-\omega_{\bf 0})/{T}}}.
\end{equation}
Although finding $F_N$ is hard in the general case, the positivity of $F_N$ follows directly from Eq.~\eqref{F_N}.
Moreover, from Eq.~\eqref{Z_N_and_F_N} it follows that $Z_N = \sum^N_{n=0}F_n$.
Thus, $Z_N$ rises monotonously with $N$.
 
\subsection{The limit of $Z_N$ as $N\to+\infty$}
Below, we assume that the limit of $Z_N$ as $N\to+\infty$ exists.
In this case, we introduce the notation $Z_\infty=\lim_{N\to+\infty}Z_N$
\begin{multline} \label{Z_infinity}
	Z_\infty
	=
	\lim_{N\to+\infty} \left[ \sum_{n=1}^N(Z_n-Z_{n-1}) + Z_0 \right]
	=
	\lim_{x\to1-0} \lim_{N\to+\infty} \left[ \sum_{n=1}^Nx^n(Z_n-Z_{n-1}) + xZ_0 \right]
	=\\=
	\lim_{x\to1-0} \lim_{N\to+\infty} \left[ \sum_{n=0}^Nx^nZ_n - x\sum_{n=0}^{N-1}x^nZ_n \right]
	=
	\lim_{x\to1-0} (1-x) \sum_{N=0}^{+\infty}x^NZ_N.
\end{multline}
Thus, we express $Z_\infty$ through the generating function of $Z_N$.

\subsection{An exact expression for $Z_N$ in 2D and in 3D for quadratic dispersion}
To find an exact expression for $Z_N$ we use its generating function (see~Eq.~\eqref{generating_function})
\begin{equation} \label{generating_function_exact}
\sum_{N=0}^{+\infty}x^NZ_N 
=
\exp \left\{  
- \sum_{\bf k} \ln \left[ 1 - x e^{-(\omega_{\bf k}-\omega_{\bf 0})/T} \right]
\right\}
\end{equation}
Below we consider the quadratic dispersion, $\omega_{\bf k} = \omega_{\bf 0} + \alpha {\bf k}^2$, in cases of 2D and 3D.

\paragraph{Polaritons in 2D.}
In 2D, the density of states is $\nu(\omega)=V/4\pi\alpha$ and Eq.~\eqref{generating_function_exact} takes the form
\begin{equation} \label{2D_generating_function}
\sum_{N=0}^{+\infty} x^NZ_N 
= 
\frac{\exp\left[ G_{2D}~{\rm Ln}_2(x) \right]}{1-x},
\end{equation}
where $G_{2D} = VT/4\pi\alpha$ is the number of states in the energy range $T$ above ${\bf k}={\bf 0}$ and ${\rm Ln}_2(x)$ is the polylogarithm of order~$2$.

In Eq.~\eqref{2D_generating_function}, we use the expansion of the exponent in the series of Bell polynomials~\cite{comtet2012advanced} and the expansion of $(1-x)^{-1}$ in a Taylor series, then we collect the coefficients with equal powers of $x$ on the left and right sides and obtain
\begin{equation} \label{2D_partition_function}
Z_N = 
\sum_{n=0}^{N} \frac{1}{n!}
B_{n}\left( \frac{1!~G_{2D}}{1^2}, \frac{2!~G_{2D}}{2^2},~...~, \frac{n!~G_{2D}}{n^2} \right),
\end{equation} 
where $B_n$ is the $n$-th complete exponential Bell polynomial~\cite{comtet2012advanced} and $B_0=1$. 
Thus, in 2D, $Z_N$ is fully determined by a dimensionless parameter $G_{2D}$ and rises monotonically from $Z_0=1$ to $Z_\infty=\exp(G_{2D}\zeta(2))$, where $\zeta(2)$ is the value of the zeta function at~$2$.

\paragraph{Polaritons in 3D.}
In 3D, the density of states is~$\nu(\omega)=(V/4\pi^2\alpha)\sqrt{(\omega-\omega_{\bf 0})/\alpha}$ and Eq.~\eqref{generating_function_exact} takes the form
\begin{equation} \label{3D_generating_function}
\sum_{N=0}^{+\infty} x^NZ_N 
= 
\frac{\exp\left[ G_{3D}~{\rm Ln}_{5/2}(x) \right]}{1-x},
\end{equation}
where $G_{3D} = V(T/4\pi\alpha)^{3/2}$ is the number of states in the energy range $T$ above ${\bf k}={\bf 0}$ and ${\rm Ln}_{5/2}(x)$ is the polylogarithm of the order $5/2$.

From Eq.~\eqref{3D_generating_function} it follows that
\begin{equation} \label{3D_partition_function}
Z_N = 
\sum_{n=0}^{N} \frac{1}{n!}
B_{n}\left( \frac{1!~G_{3D}}{1^{5/2}}, \frac{2!~G_{3D}}{2^{5/2}},~...~, \frac{n!~G_{3D}}{n^{5/2}} \right),
\end{equation} 
Thus, in 3D, $Z_N$ is fully determined by the dimensionless parameter $G_{3D}$ and rises monotonically from $Z_0=1$ to $Z_\infty=\exp(G_{3D}\zeta(5/2))$, where $\zeta(5/2)$ is the value of the zeta function at~$5/2$.

\paragraph{The calculation $Z_N$.}
We note that the direct application of Eq.~\eqref{2D_partition_function} and~\eqref{3D_partition_function} is not the fastest way to calculate $Z_N$.
This is due to the great computational difficulty of the complete Bell functions for $N\gg1$. 
The optimal strategy to obtain $Z_N$ is to use the recurrence relation, discussed in the main text.

\section{Proof of Eqs.~\eqref{2D_N=infty} and~\eqref{3D_N=infty}} \label{sec:appendix_b}
According to Eq.~\eqref{n_k_N}, when $N\gg1$,
\begin{equation}
	\frac{N - \langle \hat n_{{\bf k}={\bf 0}} \rangle}{N}
	= 
	\frac{\sum_{n=0}^{N-1} (Z_N-Z_n)}{NZ_N}
	\approx 
	\frac{\sum_{n=0}^{N-1} (Z_\infty-Z_n)}{NZ_\infty}.
\end{equation}
We consider $\sum_{n=0}^{N-1} (Z_\infty-Z_n)$ in more detail.

In 2D, we have
\begin{multline} \label{asymptotics_sum_2D}
	\lim_{N\to+\infty}\sum_{n=0}^{N-1} (Z_\infty-Z_n) =
	\lim_{N\to+\infty}\lim_{x\to 1-0}\sum_{n=0}^{N-1} (Z_\infty-Z_n)x^n = \\ =
	\lim_{x\to 1-0}\frac{\exp(G_{2D}{\rm Ln}_2(1))-\exp(G_{2D}{\rm Ln}_2(x))}{1-x} =
	Z_\infty G_{2D}\lim_{x\to 1-0}{\rm Ln}_1(x) = 
	Z_\infty G_{2D}\lim_{N\to+\infty}\ln (N) .
\end{multline}
Therefore, when $N\gg1$
\begin{equation} \label{appendix_2D_N=infty}
	\frac{N - \langle \hat n_{{\bf k}={\bf 0}} \rangle}{N} \approx \frac{G_{2D}}{N} \ln N
\end{equation}
Eq.~\eqref{2D_N=infty} directly follows from Eq.~\eqref{appendix_2D_N=infty}.

In 3D, we have
\begin{multline} \label{asymptotics_sum_3D}
	\lim_{N\to+\infty}\sum_{n=0}^{N-1} (Z_\infty-Z_n) =
	\lim_{N\to+\infty}\lim_{x\to 1-0}\sum_{n=0}^{N-1} (Z_\infty-Z_n)x^n = \\ =
	\lim_{x\to 1-0}\frac{\exp(G_{3D}{\rm Ln}_{5/2}(1))-\exp(G_{3D}{\rm Ln}_{5/2}(x))}{1-x} =
	Z_\infty G_{3D}\lim_{x\to 1-0}{\rm Ln}_{3/2}(x) = 
	Z_\infty G_{3D} \zeta(3/2).
\end{multline}
Therefore, when $N\gg1$,
\begin{equation} \label{appendix_3D_N=infty}
	\frac{N - \langle \hat n_{{\bf k}={\bf 0}} \rangle}{N} \approx \frac{G_{3D}}{N} \zeta(3/2).
\end{equation}
Eq.~\eqref{3D_N=infty} directly follows from Eq.~\eqref{appendix_3D_N=infty}.

\section{Derivation of the Eq.~\eqref{reduced_dm}} \label{sec:appendix_c}
To derive Eq.~\eqref{reduced_dm}, we denote the trace over all the states, except for the state $\bf k$, as ${\rm Tr}'_{\bf k}$.
This notation allow us to write the reduced density matrix $\hat \rho_{\bf k}$ of the state with the wave vector $\bf k$ in terms of the full density matrix $\hat \rho$ in the simple form 
\begin{equation} \label{reduced_dm_k_def}
\hat \rho_{\bf k} = {\rm Tr}'_{\bf k} \hat \rho
\end{equation}
Using the Eq.~\eqref{thermolized_dm}, we obtain
\begin{multline} \label{reduced_dm_k_zk}
\hat \rho_{\bf k} = 
\sum_{N=0}^{+\infty} P_N {\rm Tr}'_{\bf k} \hat \rho_N = 
\sum_{N=0}^{+\infty} P_N \frac{1}{Z_N} \sum_{n=0}^N | n \rangle \langle n | 
\sum_{ {\rm config} : \sum_{{\bf q}\neq{\bf k}} n_{\bf q}=N-n } 
e^{-n(\omega_{\bf k}-\omega_{\bf 0})/{T} - \sum_{{\bf q}\neq{\bf k}} n_{\bf q}(\omega_{\bf q}-\omega_{\bf 0})/{T}} =\\ =
\sum_{N=0}^{+\infty} P_N \frac{1}{Z_N} \sum_{n=0}^N | n \rangle \langle n | e^{-n(\omega_{\bf k}-\omega_{\bf 0})/{T}} z_{N-n}({\bf k}) =
\sum_{n=0}^{+\infty} | n \rangle \langle n | e^{-n(\omega_{\bf k}-\omega_{\bf 0})/{T}} \sum_{N=n}^{+\infty} P_N \frac{z_{N-n}({\bf k})}{Z_N},
\end{multline}
where $| n \rangle$ is the state, that satisfies $\hat n_{\bf k} | n \rangle = n | n \rangle$, and $z_{N}({\bf k})$ is the partition function of all the states but the state with wave vector $\bf k$
\begin{equation} \label{zk_def}
z_{N}({\bf k}) = 
\sum_{ {\rm config} : \sum_{{\bf q}\neq{\bf k}} n_{\bf q}=N } 
e^{- \sum_{{\bf q}\neq{\bf k}} n_{\bf q}(\omega_{\bf q}-\omega_{\bf 0})/{T}} 
\end{equation}
By analogy with Eq.~\eqref{generating_function}, we obtain the generating function for $z_{N}({\bf k})$
\begin{equation} \label{zk_generating_function}
\sum_{N=0}^{+\infty} x^N z_{N}({\bf k}) = 
\prod_{{\bf q}\neq{\bf k}}
\frac{1}{1 - xe^{- (\omega_{\bf q}-\omega_{\bf 0})/{T}}}
\end{equation}
Comparing~Eqs.~\eqref{generating_function}~and~\eqref{zk_generating_function}, we conclude $\sum_{N=0}^{+\infty} x^N z_{N}({\bf k}) = (1 - xe^{-(\omega_{\bf k}-\omega_{\bf 0})/{T}}) \sum_{N=0}^{+\infty} x^N Z_{N}$. 
Therefore,
\begin{equation} \label{zk_Z}
z_{N}({\bf k}) = Z_N - e^{- (\omega_{\bf k}-\omega_{\bf 0})/{T}} Z_{N-1}
\end{equation}
where we formally put $Z_{-1}=0$.
Substitution of Eq.~\eqref{zk_Z} into Eq.~\eqref{reduced_dm_k_zk} leads to Eq.~\eqref{reduced_dm}.

\end{document}